\documentclass[traditabstract]{aa} 
\usepackage{graphicx}
\usepackage{txfonts}
\usepackage{verbatim} 
\usepackage{amssymb}

\newcommand{\comments}[1]{} 

\begin{document}
\title{On the effect of image denoising on galaxy shape measurements}

\titlerunning{Image denoising and galaxy shape measurements}


\author{G. Nurbaeva\inst{1} \and F. Courbin\inst{1} \and M. Gentile\inst{1} \and
G.~Meylan\inst{1} }

 \institute{Laboratoire d'astrophysique, Ecole Polytechnique F\'ed\'erale de Lausanne
(EPFL), Observatoire de Sauverny, CH-1290 Versoix, Switzerland}

\date{Received; accepted }

\abstract{Weak gravitational lensing is a very sensitive way of measuring cosmological parameters, 
including dark energy, and of testing current theories of gravitation. In practice, this requires exquisite 
measurement of the shapes of billions of galaxies over large areas of the sky, as may be obtained with the
EUCLID and WFIRST satellites. For a given survey depth,  applying image denoising to 
the data both improves the accuracy of the shape measurements and increases the number density of
galaxies with a measurable shape. We perform simple tests of three different denoising techniques, 
using synthetic data. We propose a new and simple denoising method, 
based on wavelet decomposition of the data and a Wiener filtering of 
the resulting wavelet coefficients. When applied to the GREAT08 challenge dataset, this technique allows 
us to improve the quality factor of the measurement ($Q$; GREAT08 definition), by up to a factor of two. We demonstrate that the typical pixel size of the EUCLID optical channel will allow us to use image 
denoising. }


\maketitle
\section{Introduction}

The observed accelerated expansion of the Universe (\cite{Riess98, Perl99}) can currently be 
explained by either the existence of a repulsive force associated with so called ``dark energy'', or an erroneous
description of gravity by General Relativity on large spatial scales (for a review see 
\cite{Frieman2008}). Both explanations have profound implications for our understanding of 
cosmology and physics in general and are the main motivations of future large cosmological 
surveys. These surveys, such as the ESA EUCLID satellite project (\cite{Refregier2010}) 
combine several complementary cosmological probes to constrain the 
cosmological parameters, including the dark energy equation of state parameter $w(z)$ 
and its evolution with redshift. 

The main cosmological probe to be employed by EUCLID is weak gravitational lensing, also known as cosmic 
shear. The observational signature of 
cosmic shear is an apparent distortion of the image of distant galaxies under the influence of 
gravitational lensing by a foreground potential well. The exact way in which the galaxies are
distorted is very sensitive to the dark matter and dark energy distributions in the foreground
large scale structures, hence providing an efficient tool for cosmological measurements. 
While the first detections of cosmic shear are already a decade old and
were performed on data with relatively limited field of view and depth
(\cite{Bacon2000, Waerbeke2000, Wittman2000}), the use of cosmic shear 
in terms of cosmological applications
requires a major space survey. However, the effectiveness of the method
in constraining cosmology
relies on image processing techniques that measure the shapes of individual galaxies in
the most accurate possible way.
These techniques must provide solutions to the four following 
problems: (i) the degradations caused by the dominating
Poisson noise, (ii) the sampling adopted to represent the data, (iii) 
 the convolution by the instrumental point spread function (PSF) and its 
possible variations across the field of view, and (iv) the measurement of the cosmic shear 
itself and its power spectrum from all the galaxy shape measurements, i.e., billions of galaxies.

The techniques currently in use to measure cosmic shear are sufficient to detect it
and even sometimes to reconstruct the 3D mass map of large scale structures 
(e.g., \cite{Massey2007}) but it is estimated that a tenfold improvement in the precision of galaxy shape 
measurements is needed to place stringent constraints on cosmological models.
Thanks to both the STEP programs (\cite{Heyman, Massey})
and the GREAT08\footnote{\tt \footnotesize http://www.greatchallenges.info/} challenge (\cite{Bridle2010}), the lensing
community has made excellent progress toward meeting this goal.  
However, even the most successful shear measurement
methods see their accuracy decrease significantly under
high noise conditions. It is therefore of interest to develop suitable denoising 
techniques that are capable of solving the difficult problem of removing noise without compromising
the fragile shear signal.

In the present work, we focus on the effects of denoising and 
pixelisation on shape measurement. Both are closely connected
since the spatial frequencies contained in the galaxy images change
with the adopted pixel size, but not the noise frequency. Moreover, 
for a given exposure time, changing the pixel size affects the signal-to-noise ratio ({SNR}) of the data.

For this reason, it is important to explore the large parameter space of the problem and
to weight the relative impacts of different samplings and {SNR}s on the shear measurement.
To investigate this problem, we use sets of synthetic galaxies with known ellipticities, for 
different resolutions and samplings spanning a range of observational setups. 
Using these data, as well as a subset of the GREAT08 data (\cite{Bridle2010}), we test the 
performance of  three popular denoising algorithms: median filtering, Wiener filtering, and  discrete wavelet transform (DWT). We also propose a new denoising method based on
a combination of wavelet transform and Wiener filtering.

\begin{figure*}[t!]
\begin{center}
\includegraphics[width=\textwidth]{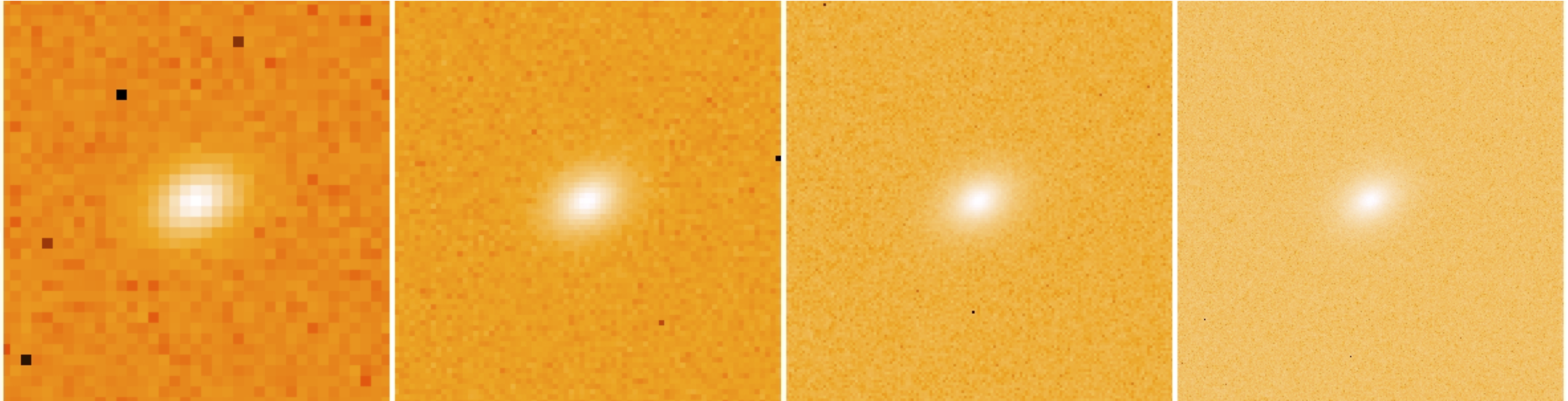}
\caption{Artificial galaxies of different resolutions. From left to right 
the stamp size is $40, 80, 160$ and $320$ pixels on a side, corresponding to 
a mean {SNR}  per pixel of $1.5, 0.75, 0.37$ and $0.19$, respectively. The
first two samplings on the left correspond to the ``realistic sampling" and the  ``small sampling" cases,
 respectively (see text).}
\label{simul}
\end{center}
\end{figure*}

\section{Methods}

\subsection{Median filter}

The median filter is a nonlinear denoising technique widely used in
digital image processing. Apart from its simplicity, median filtering
has two important properties: firstly, it is particularly
effective for images corrupted by Poisson noise and secondly, it preserves edges in images.

A median filter works by sliding a box of given size (3$\times$3 pixels in our case) 
over the image, replacing the central value by the median of its neighboring pixels (\cite{Arce,AriasDonoho})

\subsection{Wiener filter}

The Wiener filter uses a least mean squares filtering algorithm (\cite{Wiener}) based
on a stochastic framework that minimizes the mean square error in the noise. 
The Wiener filter has become a classical
signal smoothing technique and is widely used in signal processing (\cite{Khireddine,Press07}). In our study, we use the following simplified algorithm.

If we define $I$  to be the input brightness of the pixel of the
noisy image, the output brightness $I_{wiener}$ of the denoised pixel is then given by

\begin{equation}
{I_{wiener}}=
\left\{ 
\begin{array}{rr} 
I\left(1-\sigma_n^2/\sigma_w^2\right),    &   \sigma_w^2\geq\sigma_n, \\
0,     &   \sigma_w^2<\sigma_n,
\end{array}
\right. 
\label{Wiener_eq}
\end{equation}

\noindent where $\sigma_{n}$ is the estimated mean standard deviation of the noise in the image and  $\sigma_{\rm w}^2$  is an average variance in the pixel values  calculated in a local window of 3$\times$3 pixels.

The algorithm implemented in this work uses the scientific libraries (\cite{SciPy}) available with 
the Python programming language.

\subsection{Discret wavelet transform (DWT)}
\label{DWT}

\begin{figure*}[t!]
\begin{center}
\includegraphics[width=\textwidth]{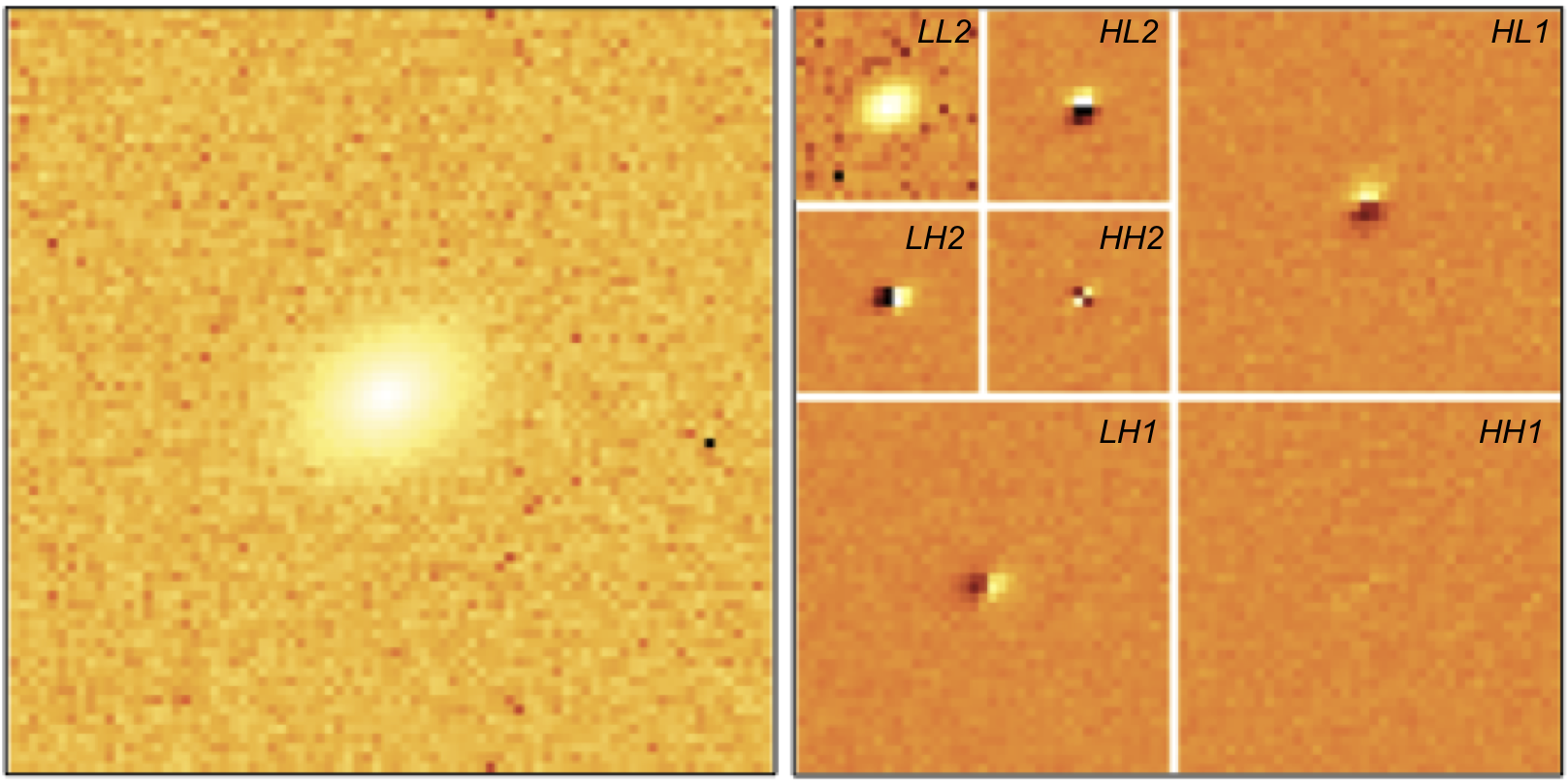}
\caption{DWT decomposition of the galaxy image, with Poisson noise. {\it Left:} original 
galaxy image.
{\it Right:} two consecutive resolution levels, as explained in Sect~\ref{DWT}.}
\label{decomposition}
\end{center}
\end{figure*}

Wavelet analysis is an efficient and fast computational technique widely used for data compression 
 and noise reduction  (\cite{Bruce1996}). In our study, we apply the 2D discrete wavelet transform (DWT). 
Denoising in wavelet space involves three steps: (i) linear forward wavelet decomposition, (ii) 
shrinkage of wavelet coefficients,  and (iii) linear inverse wavelet reconstruction.

As the basis function 
we adopt the Haar wavelet, which is orthogonal  and computationally simple.  The latter property is of primary importance to preserving shape invariance. 

  The Haar wavelet is defined by two basic functions: a scaling function $\phi$ and a wavelet function, called the mother wavelet $\psi$. The set of basic functions for the 1D case is given by

\begin{equation}
\left\{ 
\begin{array}{rr}
\phi_i^j(x)= \phi(2^{k-j}x-i)\,\,  &  \\
  \psi_i^j(x)= \psi(2^{k-j}x-i), &\,\,\,\,\,  i = 0, 1, 2, ..., N-1
\end{array}
\right. 
\label{cascade_equations}
\end{equation}

\noindent  where  ${N=2^k-j}$ is the number of wavelet coefficients, which also defines the size of the subband of a given decomposition  level $j$, where $k$ is the coarsest level.
The scaling function and the mother wavelet are defined as follows:

\begin{equation}
\phi({x})=
\left\{ 
\begin{array}{rr} 
1, & 0\leq x<1 \\ 
0, & otherwise
\end{array}
\right. 
\,\,\,\,\,\,\,\,\,\,\,\,\,\,\,
\psi({x})=
\left\{ 
\begin{array}{rr} 
1, &  0\leq x<1/2 \\ -1, & 1/2\leq x<1 \\ 0, & otherwise
\end{array}
\right. 
\end{equation}

To decompose a two-dimensional image, the coefficients are obtained by multiplying the one-dimensional scaling and the wavelet functions both in the horizontal and vertical directions. For each resolution level, 
the image is devided into four images 
of coefficients, called subbands. 
 The first, often labeled $LL$ (low-low), contains the main (low) frequency features of the signal. The three others are dominated by
the noise in the horizontal direction, $HL$ (high-low), vertical direction, $LH$ (low-high), and diagonal direction, $HH$ (high-high). Iterating the described scheme (Eq.~\ref{decomposition}), one can obtain an image sequence with a cascading structure as illustrated in Fig.~\ref{cascade_equations}.

The most important part of the DWT denoising technique is the wavelet shrinkage, which drives the efficiency of denoising. 
The wavelet shrinkage  is applied to the subbands associated with the noise: $HL$, $LH$, and $HH$. 
The classical way to suppress the noise by shrinking the wavelet coefficients is to apply a threshold to the 
wavelet coefficients. There are two types of thresholding algorithms:  soft and hard thresholding. For a given
value of the threshold $T$, hard thresholding sets all coefficients less than $T$ to zero. For the soft thresholding, $T$ is subtracted from all coefficients greater than $T$ (see e.g., \cite{Vetterli1995}).

\subsubsection{Bayes thresholding}

The  wavelet shrinkage step depends heavily on the choice of the thresholding scheme. Popular
thresholding methods include Stein's unbiased risk estimate (SURE; \cite{Dohono94}) and universal thresholding (\cite{Dohono95}), 
the latter depending on image size. A more efficient thresholding scheme is the Bayes wavelet threshold 
proposed by \cite{GraceChang}, which uses a different threshold level for each of the three $HL$, $LH$, 
and $HH$ noise subbands. This adaptive threshold, $T_{\rm Bayes}$, is computed to be

\begin{equation}
T_{\rm Bayes} = \frac{\sigma_n^2}{ \sigma_x},
\label{T_bayes1}
\end{equation}

\noindent where  $\sigma_x$ is  the
standard deviation in the noiseless coefficients in a given subband, which can be estimated as

\begin{equation}
\sigma_{\rm x} = \sqrt{max(\sigma_{\rm y}^2 - \sigma_n^2, 0)},
\label{T_bayes2}
\end{equation}

\noindent where $\sigma_{\rm y}^2$ is the variance in the coefficients in a subband. 
  If  ${\sigma_x=0}$ then $T_{\rm Bayes} $ diverges to infinity, meaning that all coefficients in the corresponding subband must be set to zero.

\subsubsection{Combined DWT-Wiener filter: a new thresholding scheme}

The DWT-Bayes thresholding gives good results 
for general-purpose imaging of relatively small dynamical range. However, it is quickly limited 
when dealing with astronomical images where the intensity levels vary sharply on spatial scales in the 
order of the PSF size. 
Standard thresholding, where $T$ is the same across each wavelet subband, degrades galaxy shapes
with steep intensity profiles, especially when images are critically sampled. 
This is simply because the light profile between two neighboring pixels is much 
steeper than the estimated noise per pixel, $\sigma_n$. The resulting denoised image is therefore too smooth
in rather flat areas such as the image background, and too noisy in areas of larger dynamical range, i.e., 
where galaxies lie. This has a significant effect on the shape measurement  
accuracy.

For this reason, we propose a simple and effective
method that combines the Haar DWT and the classical Wiener algorithms in the following way:

\begin{enumerate}

\item We decompose the image and calculate the $HL_j$, $LH_j$, and $HH_j$ subbands of wavelet coefficients for all resolution levels ${j = 1, 2, ... , k}$, where $k$ is the coarsest level.

\item We then odify the wavelet subbands $HL_j$, $LH_j$, and $HH_j$ using the kernel Wiener algorithm (Eq.~\ref{Wiener_eq})

\item We finally apply the inverse DWT to the modified wavelet coefficient to reconstruct the denoised image. 

\end{enumerate}

In other words, we apply the Wiener method to the wavelet coefficients rather than to the data pixels themselves. 
We show in the following that, for images of faint and small galaxies, this simple method not only 
conserves the shape of the galaxies, but also improves their measurement.

\begin{figure*}[t!]
\begin{center}
\includegraphics[width=\textwidth]{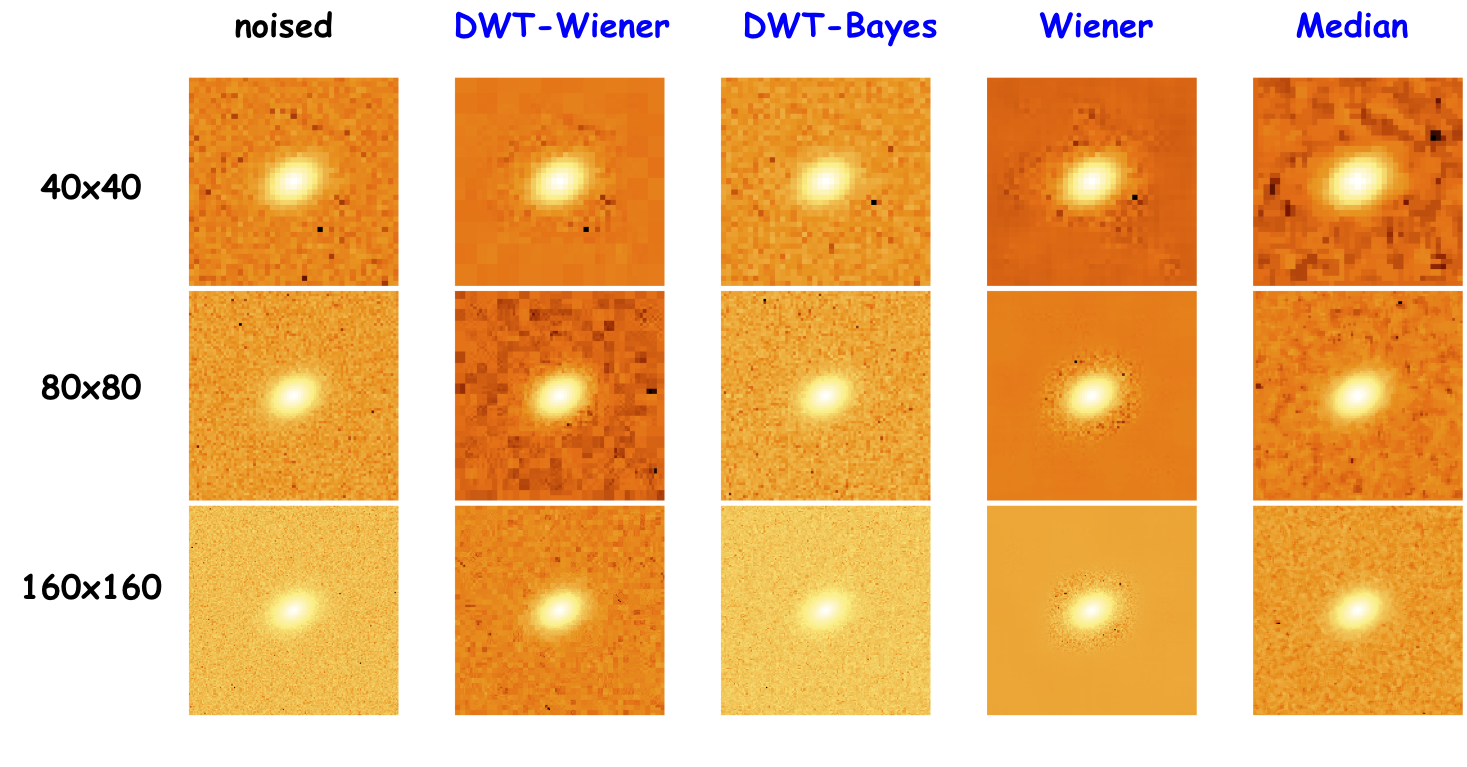}
\caption{Examples of different denoising methods, using simulated data with three different samplings.} 
\label{denoising}
\end{center}
\end{figure*}

\section{Synthetic data}
\label{data}

We use a set of 10 000 simulated galaxies in order to (i)  test the effect of the different denoising 
techniques on the shape measurement and (ii) estimate how this behavior depends on  the
sampling and the {SNR}  of the data. Each galaxy is represented by 
a Sersic profile of known ellipticity and Sersic index. This profile is sampled 
on five different grids of pixels. In doing this, we keep the size of the galaxy fixed 
on the plane of the sky. 

 Our finest sampling has galaxies with a full-width-half-maximum of 
FWHM = 17.3 pixels.  Each of the 10 000 galaxy images is represented on a stamp of
$320\times320$ pixels. We then degrade the sampling by a factor of two, four times in a row, 
to produce galaxy images with FWHM $\sim$ 17, 9, 5, 3 pixels, which correspond, respectively, to 
image sizes of  320, 160, 80, and 40 pixels on a side.

We then add Poisson noise to the simulated data, assuming that the data are sky-dominated, i.e, 
to a good approximation the amplitude of the noise is the same for all pixels across the galaxy image. 
We generate noisy images that mimic those of the GREAT08 challenge (\cite{Bridle2010}) with 
a standard deviation set to the value of $\sigma_{n}=1000$ for all pixels.  Before adding the noise, we
scale the galaxy images so that we probe a range of realistic {SNR}. 
In the rest of the paper, we refer to the mean SNR per pixel, i.e., 
\begin{equation}
SNR= \sum_{j=1}^{N} \frac{I_{j}}{N \times \sigma_n},  
\label{eq:SNR}
\end{equation}
where $I_j$ is the image value at pixel $j$ and $N$ is the total number of pixels in the
image. Since the exposure time is limited in real sky surveys, improving the sampling 
of the data is done to the cost of a lower {SNR} per pixel. All our simulated images
are computed for a fixed integration time. 

We use four different samplings, characterized by the typical FWHM of 
the simulated galaxies. The first two samplings have FWHM $\sim$ 3 and 5 pixels which are typical values for a space
mission such as  EUCLID. In the following, we refer to these as ``real sampling" data. 
We also use two smaller samplings of FWHM $\sim$ 9 and 17 pixels. We refer to these as ``small sampling" data. 
Fig.~\ref{simul} give examples of simulated images for the same realization of a galaxy.
Our simulations span the {SNR} range between 0.05 to 4.0. 

For reference,
the {SNR} of the GREAT08 challenge data range from 0.68 to 2.6 in the ``low noise" dataset
and from 0.003 to 0.38 for the high noise dataset. The FWHM of the galaxies in 
GREAT08 varies from 1.1 to 14.5 pixels.

\begin{figure*}[t!]
\begin{center}
\includegraphics[scale=0.85]{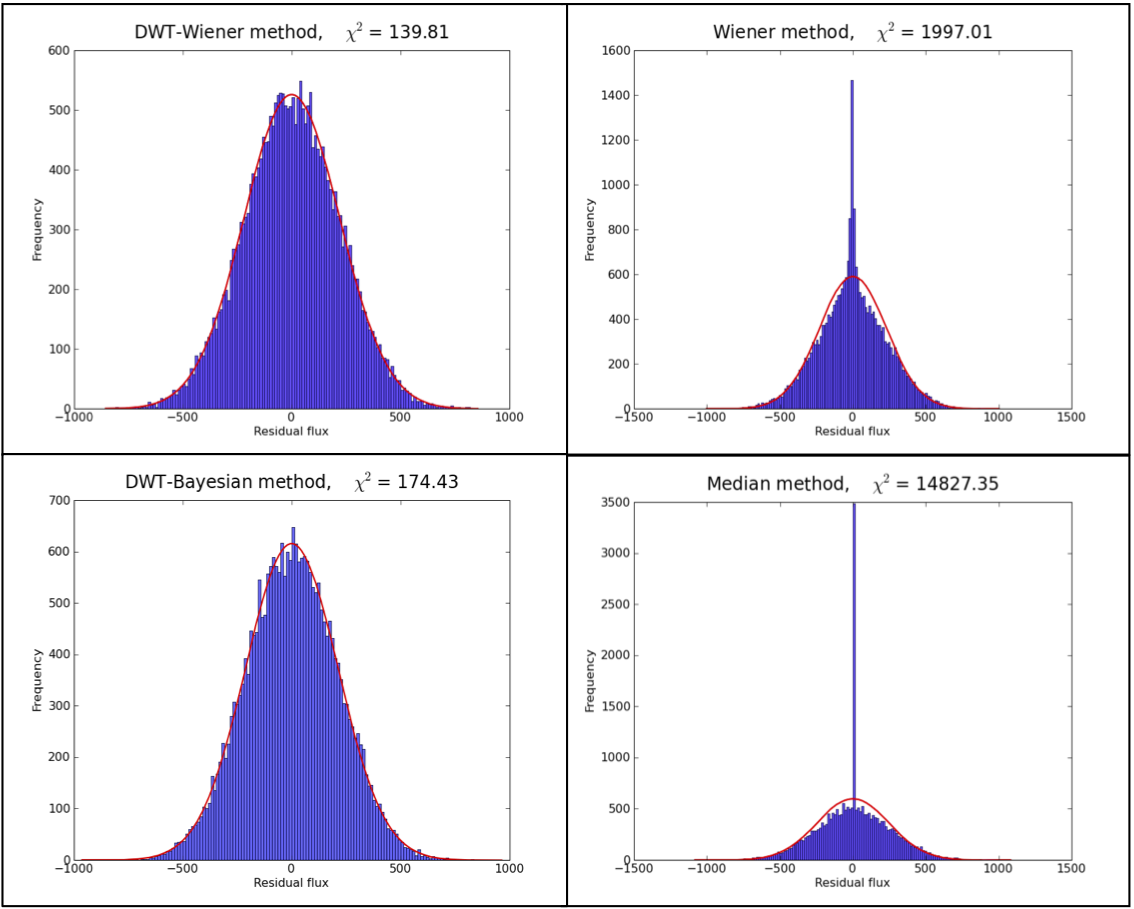}
\caption{Effect of the four denoising methods on the noise properties of the original data. 
Each panel shows the normalized histograms of residual images, i.e., the difference between the original noisy data and 
the denoised data. The red line shows the best-fit Gaussian. In each case, the $\chi^2$ of the fit is indicated.
The mean {SNR} in the image selected to compute the histogram is SNR=0.38.}
\label{resi_histo}
\end{center}
\end{figure*}

\section{Results and discussion}

\subsection{Denoising efficiency for synthetic galaxies}

Using the set of 10 000 artificial galaxies described in Sect.~\ref{data}, we test the performance of 
the four methods under different {SNR} and sampling conditions. Examples of denoised images
are shown in Fig.~\ref{denoising}, where it is immediately seen 
that the four methods behave very differently. 

The median filtering has a kernel size of 3$\times$3
pixels. As a consequence, when the sampling
changes, the spatial frequencies removed by the filter change with 
respect to the ones contained in the galaxy. When the sampling becomes coarse enough, 
the frequencies removed by the filter are close
to those contained in the galaxy. This is seen in Fig.~\ref{denoising}, where the size of the granulation in the 
noise becomes larger as the pixel size becomes larger (right column of the figure). 
For this reason, the performance of the median filtering are expected to degrade quickly in cases of 
coarse sampling.

The Wiener filter is a low-pass filter, which translates in Fig.~\ref{denoising} into a  strong ``flattening" of the sky noise
but almost no denoising of the galaxy itself.   The wavelet method is fully local and adaptive, explaining  the patchy aspect of the galaxy in Fig.~\ref{denoising}: low frequency signals are represented using a limited number of coefficients and a high frequency requires more coefficients 

Finally, we show the results of both DWT-Wiener and  DWT-Bayes methods. As explained earlier, the standard 
DWT-Bayes thresholding is not effective in removing noise from high dynamical range data such as astronomical 
images. In the particular case of galaxy images, using a fixed threshold for all the wavelet subbands tends to 
degrade the galaxy shape in areas where the difference in brightness between neighboring pixels is much 
higher than the standard deviation of the Poisson noise. In other words, the threshold is too high for the 
image background, leading to an excessive smoothing, whereas the same threshold proves too small for 
the high intensity pixels of the galaxy image. This phenomenon is clearly visible in the image denoised with
the DWT-Bayes threshold
(middle column of Fig.~\ref{denoising}), where the results of the denoising process leaves the original image almost
unchanged.
This is unfortunate because preserving the original intensity profile of the galaxy is essential for the accuracy of the measurement of its shape. Removing noise effectively while preserving the galaxy shape requires a very delicate denoising approach.
   
The DWT-Wiener method proposed in this paper removes noise according to the local gradient of luminosity, resulting in a more adaptive and local denoising process than DWT-Bayes. The advantage of DWT-Wiener in comparison to the classical Wiener filter is that DWT-Wiener is applied to high-frequency subbands coefficients only, which are those associated with noise, thus preserving the low frequency subband that contains the main features of the signal.
\begin{figure*}[t!]
\begin{center}
\includegraphics[scale=0.6]{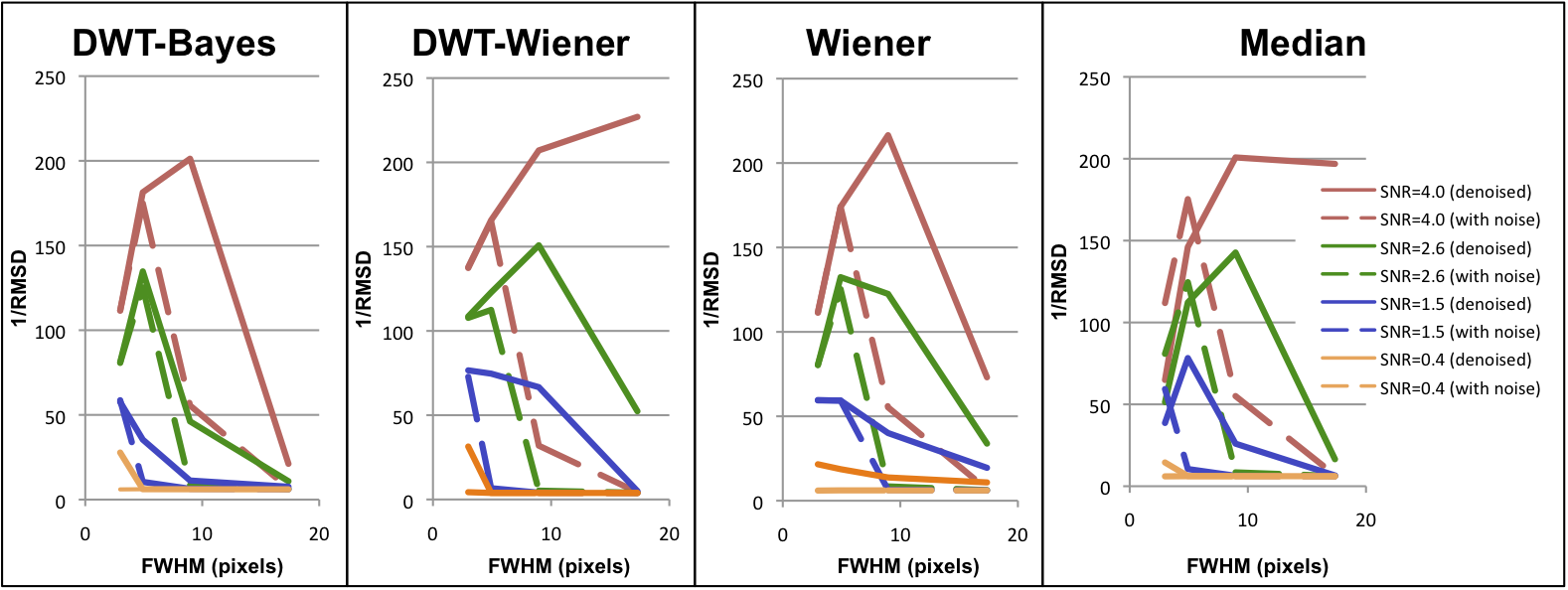}
\vskip 5pt
\includegraphics[scale=0.6]{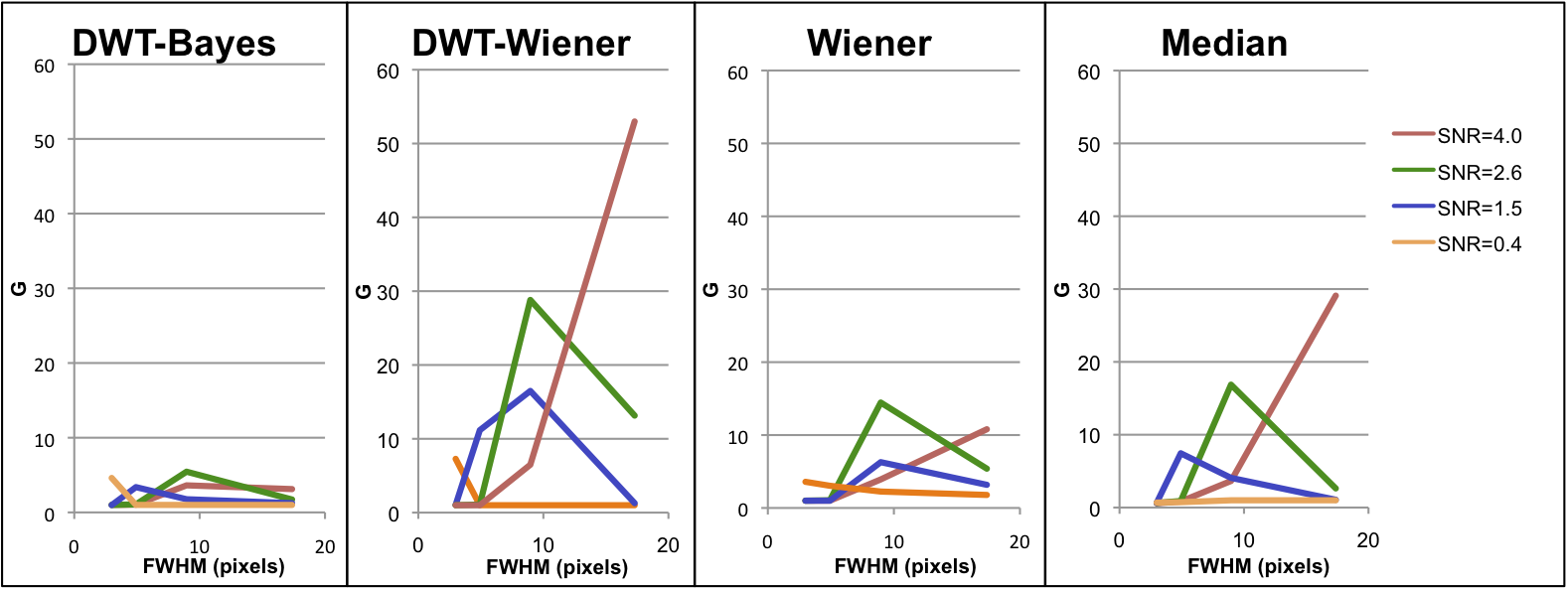}
\caption {{\it Top panels:} accuracy, $1/RMSD$, of the shape measurement using the four denoising methods 
(solid curves). The dashed curves show for comparison the results obtained with the data before denoising. 
In each panel, the color code indicates different noise levels. {\it Bottom panels:} gain ratio, G,  
for the four denoising methods. The curves are the ratio of the solid to dashed curves shown 
in the top panels.}
\label{GRplot}
\end{center}
\end{figure*}

The effect of the four methods on the noise properties of the original data 
is seen more quantitatively in Fig.~\ref{resi_histo}. In this figure,
we choose a galaxy realization with a SNR = 0.38 and sampling of {FWHM = 5} pixels (realistic sampling). 
We then compute 
the difference between the original noisy image and the denoised image, as obtained from the four methods
and plot histograms of these distributions in Fig.~\ref{resi_histo}. A method that does not affect the noise properties of the original data should ideally yield a perfectly Gaussian histogram.  Only the two local denoising methods using wavelets possess this 
important property. 

For each denoising method, we thus compute the $\chi^2$ statistics of its normalized histogram  as  (\cite{Press07})
\begin{equation}
\chi^2 =   \sum_i \frac{(N_i-n_i)^2}{n_i},  
\label{eq:chi}
\end{equation}
where $N_i$ is the number of pixels observed in the {$i$ th} bin, and $n_i$ is the number for an expected Gaussian distribution. 

Our new DWT-Wiener method has the smallest $\chi^2$, which indicates that its residuals represent the best fit to a normal distribution. We now test the performance of the four methods on the sets of 10 000 simulated galaxies described in 
Sect.~\ref{data}, without considering the effect of  convolution by the PSF.
 This is done by directly comparing the ellipticity measured for the galaxies before and after denoising.  
Using {10 000}  galaxies
ensures that an ellipticity measurement is accurate to 1\%, which is sufficient for our purposes. In this work, the galaxy ellipticity $e = e_1 + i e_2$ was estimated as

\begin{equation}
e =  {\frac{Q_{11}-Q_{22} + 2i Q_{12}}{Q_{11}+Q_{22}}},
\label{eq:ellipticity}
\end{equation}
where $Q_{mn}, \ ({m,n \in \{1,2\}})$ are the second-order quadrupole moments of the galaxy surface brightness (\cite{Bartelmann2001}).

We define the accuracy of each of the 
denoising methods as the inverse of the root mean square deviation ({RMSD})  

\begin{equation}
RMSD = \sqrt{{\frac{1}{N}}{\sum_1^N{(e_i - e_i^*)^2 }}},
\end{equation}
where $e_i$  and $e_i^*$ denote estimated and true galaxy ellipticities, respectively.

To evaluate the effectiveness of the denoising methods
in providing improved measurement over the original data, we also define a gain ratio, $G$, 
as the ratio of the shape measurement error using the noisy data, to the same quantity using the denoised data

\begin{equation}
G = \frac{RMSD_{\rm original}}{RMSD_{\rm denoised}}
\end{equation}
\begin{table}
\caption{ Gain ratio (G; see text) for the four denoising methods under different {SNR}
and sampling conditions. The {SNR} is per pixel (Eq. \ref{eq:SNR}).}
\label{GR}
\begin{tabular}{lrrrrrr}
\hline \hline
SNR & Stamp  & FWHM & Median & Wiener & DWT & DWT\\
  & size  & galaxies  &   &   &    Bayes &    Wiener\\
\hline
4.00 & 40   & 2.99 & 0.58 & 1.00 & 1.00 & 1.01\\
2.00 & 80   & 4.92 & 0.83 & 1.00 & 1.04 & 1.01\\
1.00 & 160 & 8.96 & 3.65 & 3.91 & 3.62 & 6.49\\
0.50 & 320 & 17.30 & 29.10 & 10.80 & 3.12 & 52.98\\
\hline
2.60  & 40   & 2.99 & 0.64	 & 1.00 & 1.01 & 1.01\\
1.30  & 80   & 4.92 & 0.90 & 1.06 & 1.08 & 1.10\\
0.65 & 160  & 8.96 & 16.87 & 14.47 & 5.45 & 28.81\\
0.33 & 320  & 17.30 & 2.63 & 5.40 & 1.73 & 13.15\\
\hline
1.50  & 40   & 2.99 & 0.65 & 1.00 & 0.98 & 1.05\\
0.75  & 80   & 4.92 & 7.45 & 1.00 & 3.38 & 11.17\\
0.38  & 160 & 8.96 & 4.08 & 6.28 & 1.77 & 16.48\\
0.19 & 320  & 17.30 & 1.07 & 3.18 & 1.24 & 1.28\\
\hline
0.40 & 40   & 2.99 & 0.69 & 3.59 & 2.67 & 7.28\\
0.20 & 80   & 4.92 & 0.74 & 3.05 & 0.31 & 1.01\\
0.10 & 160  & 8.96 & 1.00 & 2.25 & 0.99 & 1.00\\
0.05 & 320  & 17.30 & 1.00 & 1.79 & 1.00 & 1.00\\
\hline
\end{tabular}
\end{table}
The results are summarized in Table~\ref{GR} and Fig.~\ref{GRplot}. With no big surprise, all methods
lead to high accuracy measurements as soon as excellent sampling and high {SNR} are available. However, when  the {SNR} decreases, the performance of all methods drops sharply. This means that there is 
no reason to improve the sampling indefinitely without compensating for an increased {SNR}, i.e., exposure time.
This also means that the interest in denoising exceeds the improvement for poor  {SNR} data. Performances improve with high {SNR} data (red curves in Fig.~\ref{GRplot}, hence showing that 
denoising is as important for well exposed galaxies as for galaxies barely measurable in the original data. 

In coarse sampling conditions, the median filter
degrades shape measurement instead of improving it. This is illustrated in {Fig.~\ref{GRplot}} by the 
dashed curves systematically being above the solid ones. In Table.~\ref{GR}, the gain ratio factor, $\xi$, is
indeed lower than 1. This is because the frequencies removed by the median filter cannot be
easily controlled and depend on the size of the objects with respect to the size of the median kernel. This makes
 the use of the median filter very hazardous in general, in spite of its good performance in small
sampling conditions. A similar trend is found for the Wiener filter in realistic sampling conditions, although it is not 
so pronounced. The best gain factors are usually achieved with the DWT-Wiener method. 

Finally, for each method  there exists an optimal sampling where the gain factor is maximum. For realistic 
sampling and {SNR} conditions (blue and orange lines in Fig.~\ref{GRplot}, this sampling is about FWHM $\sim$ 9 
pixels. By realistic conditions, we mean conditions similar to those in the GREAT08 challenge, i.e.,  
data quality mimicking that of the EUCLID satellite in terms of sampling and {SNR}.

\subsection{Denoising of the GREAT08 challenge dataset}

\begin{table}
\caption{$Q$ factor for KSB deconvolution for 15 galaxy sets of GREAT08 Low Noise-Blind (LNBL) and 100 sets of GREAT08 Real Noise-Blind (RNBL)}
\begin{tabular}{lrrrr}
\hline \hline
 Dataset & $\alpha$  & original & DWT-Bayes & DWT-Wiener \\
  &  &  &    denoising &    denoising \\
\hline
                      & 0.01  & 32.37 & 32.29 & 32.37 \\
                      & 0.05  & 32.37 & 32.36 & 32.79 \\
LNBL            & 0.10  & 32.37 & 32.37 & 61.53 \\
(Low-Noise) & 0.50 & 32.37 & 32.21 & 60.13 \\
                       & 1.00  & 32.37 & 31.60 & 75.70 \\
\hline
                          & 0.01  & 11.54 & 11.32 & 11.51 \\
                          & 0.05  & 11.54 & 11.32 & 11.54 \\
RNBL               & 0.10  & 11.54 & 11.55 & 15.94 \\
(Real-Noise)    & 0.50 & 11.54 & 11.55 & 15.12 \\
                          & 1.00 & 11.54 & 3.34 & 4.67 \\
\hline
\end{tabular}
\label{GREAT08_KSB}
\end{table}

The goal of the previous section was to test the effect of both denoising and sampling on the quality
of the shape measurement of galaxies, in the absence of other numerical or instrumental disturbances. 
For this reason, the convolution of the galaxy images by the instrumental PSF is not included.

We now test the two most effective denoising methods, i.e., the DWT-Bayes and the DWT-Wiener, 
against the additional effect of PSF deconvolution and centroid shifts. This 
requires the use of  a shape measurement method. For the sake of simplicity, we choose the
KSB algorithm (\cite{KSB}) with the code developed by Catherine Heymans and Ludovic Van Waerbeke (\cite{Heyman}), which is widely used, public, and efficient in terms of computing time. 
In addition, it does not rely on any fit of an arbitrary galaxy profile to the data. 

The GREAT08 challenge dataset is ideal for carrying out our test in both low noise and real noise conditions. 
For this purpose, we use the whole Low Noise-Blind (LNBL) dataset and a subset of 100 frames of the Real 
Noise-Blind (RNBL) dataset (see the GREAT08 handbook for more detail, \cite{Bridle2009}). The total number of galaxies used is therefore 15 000 in low noise and 
100 000 in real noise conditions. 

The PSF convolution, which modifies the effect of the noise on the galaxy shape measurement, makes it
necessary to control the effect of denoising, prior to the shape measurements itself. We therefore introduce a 
``denoising strength" that allows us to fine-tune the amount of denoising of the data. In practice, we  
replace the noise mean standard deviation, $\sigma_n$, in Eqs. (\ref{Wiener_eq}), (\ref{T_bayes1}), and (\ref{T_bayes2}),
with a fractional mean standard deviation $\alpha \sigma_n$, where  ${0<\alpha \leq1.0}$.

We run the KSB method on the denoised GREAT08 data using three 
denoising strengths $\alpha = \{0.01, 0.05, 0.1, 0.5, 1.0\}$. The results are summarized in Table.~\ref{GREAT08_KSB}, 
where the quality factor, $Q$, is defined as in Bridle et al. (2010). High values of $Q$ indicate good shape
measurements. As a sanity check,  we run the KSB algorithm on the original noisy data as well and check that
we obtain the same quality factors as in Bridle et al. (2010), i.e., $Q=32.37$ in low noise and $Q=11.54$ in real noise
conditions.

The main result  is that the DWT-Bayes method does not improve the quality factor and even degrades
the shape measurement if full denoising ${(\alpha=1.0)}$ is used. The DWT-Wiener method improves the $Q$ factor 
by a factor of almost two in low noise conditions and by 35\% in real noise conditions. Interestingly, we note that 
full denoising may degrade the results. One may Instead apply partial denoising 
($\alpha=0.1$), which allows us to achieve a significant gain in quality factor without any significant risk
of corrupting the data. 

\section{Conclusions}

We have tested four different image denoising techniques on synthetic data of faint and small
galaxies and we evaluate their effect on the shape measurement of galaxies in view of weak 
lensing studies. We have compared the performance of the algorithms for a range of {SNR} and 
sampling conditions. 

We found that simple median and Wiener filtering degrades the quality of the galaxy shape measurement 
unless very fine sampling is used. Local denoising methods such as wavelet filtering (DWT-Bayes) preserve the
shape of galaxies in fine sampling condition but not for coarser sampling. However, a simple modification 
of the thresholding scheme of the wavelet method allows us to improve the  {SNR} of the data and the 
quality of the shape measurement. This new method, DWT-Wiener, consists of applying Wiener filtering to 
the wavelet coefficients rather than to the data themselves. 

The DWT-Wiener method is tested on the GREAT08 challenge data in low noise and real noise conditions,
showing an improvement of up to a factor of two (on the quality factor $Q$) over the shape measurement using 
the original, noisy, data. 

Finally, we have shown that for a fixed {SNR} there exist an optimal sampling of the galaxy images. For the typical
{SNR} expected in weak lensing space surveys, this sampling is about FWHM $\sim$ 9 pixels. Satellites such
as EUCLID or WFIRST will have a typical pixel scale of 0.1\arcsec, allowing us to observe typical ${z=1-2}$
galaxies with almost this optimal sampling. This lends considerable hope to significantly improving future 
weak lensing measurements with image denoising.

\begin{acknowledgements}
The authors would like to thank C. Heyman and L. van Waerbeke for their KSB method code.  The authors are also grateful to Malte Tewes for fruitful discussion. This work is supported by the Swiss National Science Foundation (SNSF). 
\end{acknowledgements}

\end{document}